\documentclass[twoside]{article}
\usepackage[sc]{mathpazo}
\usepackage[T1]{fontenc}
\usepackage{microtype}
\usepackage[hmarginratio=1:1,top=32mm,columnsep=20pt]{geometry}
\usepackage{multicol}
\usepackage[hang, small,labelfont=bf,up,textfont=it,up]{caption}
\usepackage{booktabs}
\usepackage{float}
\usepackage{hyperref}
\usepackage{amsmath}
\usepackage{authblk}
\usepackage{lettrine}
\usepackage{paralist}
\usepackage{pgfplots}
\usepackage{abstract}
\usepackage{titlesec}
\usepackage{natbib}

\bibpunct{(}{)}{;}{a}{}{;}

\usetikzlibrary{positioning}
\usetikzlibrary{fit}
\usetikzlibrary{backgrounds}
\usetikzlibrary{calc}
\usetikzlibrary{shapes}
\usetikzlibrary{mindmap}
\usetikzlibrary{decorations.text}
\usetikzlibrary{shapes,arrows}

\renewcommand\thesection{\arabic{section}}
\renewcommand\thesubsection{\Alph{subsection}}

\titleformat{\section}[block]{\large\scshape\centering}{\thesection.}{1em}{}
\titleformat{\subsection}[block]{\large}{\thesubsection.}{1em}{}

\title{\vspace{-15mm}\fontsize{13pt}{10pt}\selectfont\textbf{Can Agent-Based Models Probe Market Microstructure?}}

\author[1,2]{D. F. Platt \thanks{Corresponding author, dfplatt@mweb.co.za}}
\author[1,2,3]{T. J. Gebbie}
\affil[1]{\small School of Computer Science and Applied Mathematics, University of the Witwatersrand, Johannesburg}
\affil[2]{\small QuERILab - Quantifying Emergence, Risk and Information}
\affil[3]{\small Department of Statistical Sciences, University of Cape Town, Cape Town}

\date{}

\begin{document}

\maketitle


\vspace{-1cm}

\begin{abstract}

\noindent We extend prior evidence that naively using intraday agent-based models that involve realistic order-matching processes for modeling continuous-time double auction markets seems to fail to be able to provide a robust link between data and many model parameters, even when these models are able to reproduce a number of well-known stylized facts of return time series. We demonstrate that while the parameters of intraday agent-based models rooted in market microstructure can be meaningfully calibrated, those exclusively related to agent behaviors and incentives remain problematic. This could simply be a failure of the calibration techniques used but we argue that the observed parameter degeneracies are most likely a consequence of the realistic matching processes employed in these models. This suggests that alternative approaches to linking data, phenomenology and market structure may be necessary and that it is conceivable that one could construct a useful model that does not directly depend on the nuances of agent behaviors, even when it is known that the real agents engage in complex behaviors.

\vspace{0.3cm}


\vspace{0.3cm}


\end{abstract}


\begin{multicols}{2}

\section{Introduction}

Financial agent-based models (ABMs) have seen increasing prevalence in quantitative finance literature in recent years, with a wide variety of models emerging that are capable of replicating the stylized facts of financial return time series, such as a fat-tailed distribution and volatility clustering \citep{Barde16}. This is largely due to their replacement of empirically inconsistent economic assumptions, such as that of a Gaussian return distribution and the efficient market hypothesis, with assumptions rooted in agent behaviors \citep{LeBaron05}. 

Despite this, there still exists significant skepticism as to the value and validity of this class of models, particularly in relation to current validation techniques \citep{Hamill16}. The vast majority of financial ABMs, particularly those replicating continuous double auction markets at an intraday time scale, are still validated by demonstrating an ability to produce return time series replicating a number of empirically-observed stylized facts \citep{Panayi13}. A detailed survey of such stylized facts is presented by \citet{Cont01}. Though a solid starting point, this method of validation has brought with it a number of significant challenges.

A key concern is the fact that a very large number of models with very different design philosophies are able to replicate the stylized facts of return time series equally well, leading to difficulty in model comparisons and the precise identification of which mechanisms lead to the aforementioned stylized facts \citep{Barde16}. 

This, combined with concerns that increases in the mechanistic complexity of intraday ABMs may not provide improved representations of the dynamics and processes that govern market behavior and structure on shorter time scales necessitates a more comprehensive investigation into model validation.

This leads to a need to calibrate such models to transaction data in an attempt to obtain parameters that allow the models in question to generate financial return time series with moments and other statistical characteristics comparable to empirical measurements, such that an empirically measured series and a simulated series can then be said to come from the same distribution \citep{Fabretti13}. Such investigations have been done in the work of \cite{Fabretti13} and \cite{Gilli03}, among various other studies. An exhaustive survey is presented by \citet{Kukacka16}.

The majority of these prior investigations have focused primarily on the calibration of very simple models that produce time series appropriate at a daily time scale and typically make use of closed-form approximations to calculate market prices, in the vein of closing auctions occurring at the end of each trading day in real financial markets. Models of this class include the \citet{Farmer02} and \citet{Kirman91} models. Overall, these experiments have proven relatively successful, with calibration using these methods generating satisfactory results. 

Despite this, the methods employed by \citet{Fabretti13} and \citet{Gilli03} have not been readily applied to intraday models employing realistic order matching processes approximating continuous double auction markets, with the general calibration literature for this class of models also being relatively sparse.

The underlying questions of when, and to what extent, one can use aggregate properties of financial fluctuations or return time series phenomena to probe the complex hidden features and structures inherent in real financial markets remains open. ABMs have been used to successfully extract hidden structure for aggregate return data on daily sampled frequencies \citep{Fabretti13}, but this still remains largely unexplored territory in the case of intraday data.

For this reason, we had previously applied the calibration framework of \citet{Fabretti13} in \citet{Platt16}, attempting to calibrate the \citet{JacobLeal15} model, a model representing both high- and low-frequency trader interactions in the context of a continuous double auction market. This model has a variety of features that are common in more complex intraday models, such as matching orders mechanistically rather than market clearing at or near an equilibrium price. Other examples of this class of models include the \citet{Chiarella02}, \citet{Chiarella09} and \citet{Preis06} models.

In this prior investigation \citep{Platt16}, we demonstrated and verified the ability of the model to generate well-known return time series stylized facts, while still suffering from parameter degeneracies when subjected to the established calibration procedures of \citet{Fabretti13}. We also found the dynamics of the simulated price time series to be dominated by the underlying dynamics of order prices within the model, with parameters relating to order price determination being the only meaningfully calibrated parameters. This suggested that some parameters may not have the meaning or effect that the model designer had originally intended.

This can be regarded as problematic if such agent-based modeling research focuses on the varying of parameter values to determine the effect of various phenomena in financial markets when the link between the parameters and real data from the markets being considered is weak. If these types of models are used to make both regulatory and structural inferences about market microstructure then the robustness of the ability to link parameters to real data and the ability to understand how parameters drive model behavior is essential.

The original investigation of \citet{JacobLeal15} attempts to determine the effect of different numbers of high-frequency traders on the prevalence of flash crashes in a simulated market by varying a suitable parameter. Our prior investigation found that the parameter in question in the model of \citet{JacobLeal15} was degenerate and had a poorly behaved effect on a number of moment and test statistic values of the generated price time series; this included the mean, standard deviation, kurtosis and generalized Hurst exponent \citep{Platt16}. This is also in spite of the model's ability to replicate a number of well-known stylized facts.

The source of these degeneracies is difficult to determine. They may emerge because the \citet{JacobLeal15} model generates order prices for most of its traders without consulting the current state of the limit order book (LOB), and instead uses a geometric random walk that references previous market prices. They may also be a consequence of the use of realistic order matching processes rather than closed-form approximations to market prices, which may affect the link between the underlying parameters and empirical measurements. Alternatively, it may be that the modeling framework is fundamentally incomplete and that key drivers of measured price dynamics are absent from the models themselves.

This ultimately leads us to consider two key questions in the following investigation:

Firstly, since some success was found in calibrating parameters related to order price dynamics in our previous investigation \citep{Platt16}, is it possible that despite the existence of degeneracies in parameters relating to agent behaviors, parameters which relate to the order book and market microstructure can be meaningfully calibrated?

Secondly, can aspects of market microstructure not explicitly modeled in an ABM, such as order flow correlation, be detected through calibration? In other words, can we probe model incompleteness through calibration?

We attempt to shed light on these concerns by applying the calibration methodology previously considered \citep{Platt16} to another well understood intraday ABM approximating a continuous double auction market using realistic matching procedures, but now using an order placement mechanism that is more closely related to the current state of the order book, namely the \cite{Preis06} model.

An important precedent in the literature is that relating to the zero-intelligence agent argument of \citet{Farmer05}, which suggests that very basic models with high parsimony and clear falsifiability can match observations surprisingly well. These types of models are driven by simple relationships between order flow and price process parameters with no explicit need for parameters related to more sophisticated agent behaviors. Our study therefore considers a more sophisticated model that has similar basic parameters as an important subset.

\section{The Preis et al. Model \label{Model}}

\subsection{Model Overview}

We consider the model first presented by \citet{Preis06} and later elaborated upon by \citet{Preis07}.

Referring to Figure \ref{ModelOverview}, the model consists of two agent types, {\it liquidity providers}, who submit limit orders to buy or sell an asset to a LOB, and {\it liquidity takers}, who submit market orders to buy or sell an asset to the same LOB, through a series of $T$ Monte Carlo steps. 

\tikzstyle{trader} = [rectangle, draw, fill=red!20, text width=5em, text centered, rounded corners, minimum height=4em]
\tikzstyle{process} = [rectangle, draw, fill=green!20, text width=5em, text centered, rounded corners, minimum height=4em]
\tikzstyle{terminal} = [circle, draw, fill=green!20, text width=5em, text centered, rounded corners, minimum height=4em]
\tikzstyle{line} = [draw, -latex']

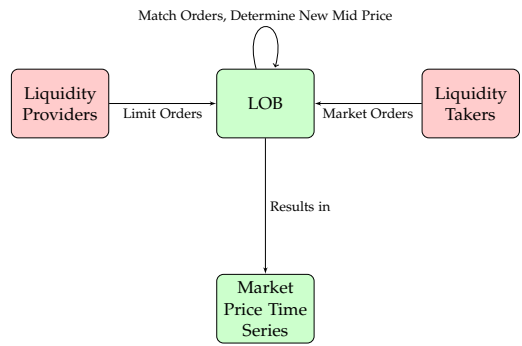
\begin{figure}[H]

\begin{center}

\scalebox{0.65}{
\begin{tikzpicture}[node distance = 4.2cm, auto] 
\node [process] (LOB) {LOB};
\node [trader, right of=LOB] (taker) {Liquidity Takers};
\node [trader, left of=LOB] (provider) {Liquidity Providers};
\node [process, below of = LOB] (price) {Market Price Time Series};
\path [line, align=left] (taker) -- node [midway, below, scale=0.8] {Market Orders} (LOB);
\path [line, align=left] (provider) -- node [midway, below, scale=0.8] {Limit Orders} (LOB);
\path [line] (LOB) edge[loop above] node [midway, above, scale=0.8] {Match Orders, Determine New Mid Price} ();
\path [line, align=left] (LOB) -- node [midway, right, scale=0.8] {Results in} (price);
\end{tikzpicture}
}

\end{center}

\caption{An illustration of the agent types and interactions within the model \label{ModelOverview}}

\end{figure}

During each of these Monte Carlo steps, active liquidity providers first submit limit orders, referencing a decision based on the current state of the LOB, after which liquidity takers submit market orders, resulting in a number of trades. 

Based on the mid price\footnote{The average of the best bid (buy order with the highest price) and best ask (sell order with the lowest price) in the LOB.} of orders in the LOB following the execution of all trades in a Monte Carlo step, a new market price for the asset can be determined, ultimately resulting in a series of market prices, one for each Monte Carlo step.

\tikzstyle{trader} = [rectangle, draw, fill=red!20, text width=5em, text centered, rounded corners, minimum height=4em]
\tikzstyle{process} = [rectangle, draw, fill=green!20, text width=5em, text centered, rounded corners, minimum height=4em]
\tikzstyle{terminal} = [circle, draw, fill=green!20, text width=5em, text centered, rounded corners, minimum height=4em]
\tikzstyle{line} = [draw, -latex']

\begin{figure}[H]

\begin{center}

\scalebox{0.5}{
\begin{tikzpicture}[node distance = 3.8cm, auto] 
\node [terminal] (start) {Simulation Start};
\node [trader, right of=start] (initialize) {Model Initialization, 10 MC Steps};
\node [trader, below of=start] (taker) {Liquidity Takers Place Market Orders};
\node [trader, left of=taker] (provider) {Liquidity Providers Place Limit Orders};
\node [trader, right of= taker] (cancel) {Liquidity Providers Cancel Limit Orders};
\node [process, below of= cancel] (midprice) {Determine New Mid Price};
\node [trader, left of= midprice] (probability) {Update Liquidity Taker Buy Probability};
\node [trader, left of= probability] (depth) {Update Liquidity Provider Placement Depth};
\node [terminal, below of=depth] (end) {Simulation End};
\path [line] (provider) -- (taker);
\path [line] (taker) -- (cancel);
\path [line] (cancel) -- (midprice);
\path [line] (midprice) -- (probability);
\path [line] (probability) -- (depth);
\path [line] (start) -- (initialize);
\path [line, align=left] (initialize) -- node [midway, right, scale=0.8] {Enter Standard MC Steps} (provider);
\path [line, align=left] (depth) -- node [midway, right, scale=0.8] {MC Step 1 \\ to $T - 1$} (provider);
\path [line, align=left] (depth) -- node [midway, right, scale=0.8] {MC Step T, Exit Standard MC Steps} (end);
\end{tikzpicture}
}

\end{center}

\caption{Flowchart describing a typical simulation of T Monte Carlo steps \label{ModelEvents}}

\end{figure}
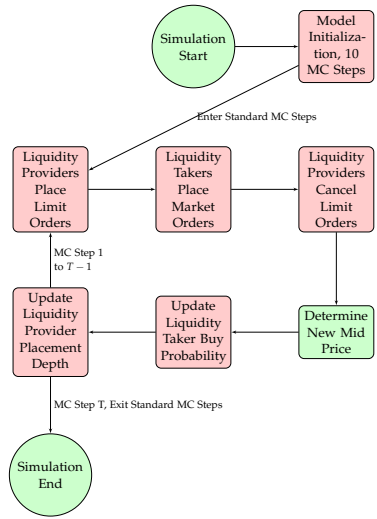

We present an outline of the overall simulation in Figure \ref{ModelEvents}, where the steps indicated are elaborated upon in subsequent subsections.

\subsection{Liquidity Provider Agent Specification}

Each simulation contains $N_A$ liquidity providers, who submit limit orders to the LOB at a given frequency, $\alpha$. This means that for any given Monte Carlo step, approximately $\lfloor \alpha N_A \rfloor$ limit orders are placed by liquidity providers. Each of the aforementioned orders is of size $1$ and is a buy order with probability $q_{provider} = \frac{1}{2}$.

The price of each of these orders is determined by the current placement depth parameter, $\lambda(t)$, and the current state of the order book, namely the current best bid, $p_b$, and the current best ask, $p_a$, where the price of a limit buy order is given by
\begin{equation}
p = p_a - 1 - \eta 
\end{equation}
and the price of a limit sell order is given by
\begin{equation}
p = p_b + 1 + \eta 
\end{equation}
where $\eta$ is an exponentially distributed random number, generated according to
\begin{equation}
\eta = \lfloor -\lambda(t)\ln(u) \rfloor
\end{equation}
and $u \sim U(0, 1)$.

The placement depth parameter is time varying, and is directly related to the buy probability of liquidity takers, $q_{taker}(t)$, discussed in more detail in subsequent subsections, and the initial placement depth parameter, $\lambda_0$. It is given by
\begin{equation}
\lambda(t) = \lambda_0 \left( 1 + \frac{\left| q_{taker}(t) - \frac{1}{2} \right|}{\sqrt{\left \langle \left[q_{taker}(t) - \frac{1}{2} \right] ^ 2 \right \rangle}}C_\lambda \right)
\end{equation}
where $C_\lambda$ is an integer parameter and $\langle [q_{taker} - \frac{1}{2}] ^ 2 \rangle$ is determined by iterating $q_{taker}(t)$ for $10 ^ 5$ Monte Carlo steps separately before the main simulation and obtaining the mean value of $[ q_{taker} - \frac{1}{2}] ^ 2$.

It should be noted that the above order placement mechanism is entirely driven by the current characteristics of the LOB. This is in contrast to the placement mechanism used by \citet{JacobLeal15}, which is based on the previous history of market prices. A more detailed LOB-based order placement mechanism is presented by \citet{Mandes15}.

During each Monte Carlo step, liquidity providers may also cancel previously placed orders. This is implemented by randomly cancelling each limit order in the LOB with probability $\delta$.

Finally, the model requires some form of initialization to produce a viable order book before actual trading can begin. This is done in a series of $10$ initial Monte Carlo steps, in which liquidity providers place limit orders around an initial price, $p_a = p_b = p_0$, with no liquidity taker activity present. This corresponds to the placement of approximately $\lfloor 10 \alpha N_A \rfloor$ limit orders.

\subsection{Liquidity Taker Agent Specification}

Each simulation contains $N_A$ liquidity takers, who submit market orders to the LOB at a given frequency, $\mu$. This means that for any given Monte Carlo step, approximately $\lfloor \mu N_A \rfloor$ market orders are placed by liquidity takers. Each of the aforementioned orders is of size $1$ and is a buy order with probability $q_{taker}(t)$, with $q_{taker}(0) = \frac{1}{2}$.

A sell market order simply executes and results in the removal of the best bid from the LOB and a buy market order simply executes and results in the removal of the best ask from the LOB, since all orders are of size $1$.

Unlike $q_{provider}$, $q_{taker}(t)$ is not fixed, but rather evolves over time. It is implemented as a mean-reverting random walk, with mean $q_{taker}(0)$, increment $\Delta_S$, and a mean reversion probability of $\frac{1}{2} + | q_{taker}(t) - \frac{1}{2} |$.

\section{Calibration Experiment Design}

\subsection{Data}

The dataset used in all calibration experiments is acquired in the TRTH \citep{Reuters16} format, presenting a tick-by-tick series of trades, quotes and auction quotes. 

We convert the dataset to a series of one-minute price bars, with each price corresponding to the final quote mid price for each minute, where the mid price of a quote is given as the average of the level 1 bid price and level 1 ask price associated with that quote, as was previously done in \citet{Platt16}. From this series of prices, we may obtain a series of log prices \footnote{This is a technical requirement of the method of \citet{Fabretti13} and is discussed in detail in \citet{Platt16}.}, which is the series we attempt to calibrate the model to.

The transaction dataset often presents events occurring outside of standard trading hours, 9:00 to 17:00, but we consider only quotes with a timestamp occurring in the period from 9:10 to 16:50 on any particular trading day. This is a result of the fact that the opening auction occurring from 8:30 to 9:00 tends to produce erroneous data during the first 10 minutes of continuous trading and the fact that the period from 16:50 to 17:00 represents a closing auction. 

In all calibration experiments, we investigate a one-week period, corresponding to a total of 2300 one-minute price bars, representing 460 minutes of trading each day from Monday to Friday.

Finally, we consider a single, liquid stock listed on the Johannesburg Stock Exchange, Anglo American PLC, over the period beginning at 9:10 on 1 November 2013 and ending at 16:50 on 5 November 2013. This dataset was also considered in \citet{Platt16}.

\subsection{Calibration Framework}

As previously discussed, we apply the calibration framework described by \citet{Fabretti13} to an intraday ABM approximating a continuous double auction market through the use of realistic order matching procedures, as opposed to a model operating at a daily time scale and using closed-form approximations to market prices.

We make use of the method of simulated moments \citep{Winker07} to construct an objective function that measures errors relating to the mean, standard deviation, kurtosis, Kolmogorov-Smirnov (KS) test and generalized Hurst exponent when comparing a log price time series measured from the data and a log price time series simulated using the \citet{Preis06} model.

We aim to minimize this objective function by employing the Nelder-Mead simplex algorithm combined with the threshold accepting heuristic \citep{Gilli03}.

We reproduce the implementation we previously described in \citet{Platt16} without any alteration, again making use of $5$ Monte Carlo replications, due to computational constraints \footnote{A single calibration experiment involving the Nelder-Mead simplex algorithm may take in excess of 7 hours when parallelized over the $2 \times 16 = 32$ workers of an HP Z840 workstation with $2$ Intel Xeon E5-2683 CPUs, making more Monte Carlo replications computationally too expensive relative to the reduction in the variance of the estimates.}.

\section{Calibration Results \label{CalResults}}

\subsection{Free and Fixed Parameters}

In all calibration experiments, we consider $6$ free parameters, with only $N_A$ set to be fixed. This particular simplification was made for the purposes of maintaining computational tractability, since increasing the value of $N_A$ can have a detrimental impact on the speed of simulation. We have chosen $N_A = 250$, as was selected by \citet{Preis06}. 

The remaining model parameters, namely $\delta$, $\lambda_0$, $C_\lambda$, $\Delta_S$, $\alpha$ and $\mu$, defined in Section \ref{Model}, are set to be free parameters and we randomly generate initial values for these parameters during each calibration experiment.

We initialize $\delta$, $\Delta_S$ and $\mu$ between $0$ and $0.1$, $\alpha$ between $0.1$ and $0.5$, $\lambda_0$ between $0$ and $200$, and $C_\lambda$ between $0$ and $20$.

\subsection{Nelder-Mead Simplex Algorithm \label{NM}}

Since we consider $n = 6$ free parameters in our calibration experiments, we begin with $n + 1 = 7$ simplex vertices, each consisting of randomly generated initial values for the $6$ free parameters.

In our experiments, it was noted that $100$ iterations of the Nelder-Mead simplex algorithm with threshold accepting almost always produced convergent behavior and we thus conducted $20$ calibration experiments with this fixed number of iterations. 

Despite the presence of convergent behavior, we find that the obtained parameter sets are somewhat different for various calibration experiments, with dependence on the set of initial vertices. We also observed this phenomenon in \citet{Platt16}, but find a greater average reduction in the initial search space when considering the parameter confidence intervals \footnote{The intervals are calculated as: $\bar{x} \pm t^{*} \frac{s}{\sqrt{n}}$, where $\bar{x}$ is the sample mean, $s$ is the sample standard deviation, $n$ is the sample size, and $t^{*}$ is the appropriate critical value for the $t$ distribution.} obtained for the \citet{Preis06} model in comparison to those obtained for the \citet{JacobLeal15} model. These confidence intervals are shown in Table \ref{NMResults}. 

\begin{table}[H]
\caption{Nelder-Mead Simplex Algorithm Calibration Results \label{NMResults}}
\begin{tabular*}{\linewidth}{@{\extracolsep{\fill}}ccc}
\hline
Parameter & 95\% Conf Int & $\frac{s}{\sqrt{n}}$ \\ \hline
$\delta$ & $[0.0554, 0.0782]$ & $0.0054$ \\
$\lambda_0$ & $[137.3818, 190.9182]$ & $12.7892$ \\
$C_\lambda$ & $[11.0808, 21.3192]$ & $2.4458$ \\
$\Delta_S$ & $[0.0272, 0.0572]$ & $0.0072$ \\
$\alpha$ & $[0.1545, 0.2517]$ & $0.0232$ \\
$\mu$ & $[0.0692, 0.1014]$ & $0.0077$ \\ \hline
\end{tabular*}
\vspace{0.2cm}
\caption*{95\% confidence intervals for the set of free parameters, obtained from 20 independent calibration experiments involving the Nelder-Mead simplex algorithm combined with the threshold accepting heuristic}
\end{table}

\vspace{-0.5cm}

This is indicative of convergence to local minima, even with the inclusion of methods aiming to overcome this problem, namely the threshold accepting heuristic. It would therefore appear that there exists significant difficulty in identifying a unique, optimal parameter set.

As was discussed by \citet{Fabretti13}, we also find that these different parameter sets form a region of feasible parameters, all producing similar objective function values, despite differences in the parameter values themselves. In \citet{Platt16}, we found that such a trend existed, but not as prominently as in this case, where we now find that all the calibration experiments conform to this behavior. 

It is also worth noting that, as was found in \citet{Platt16}, the majority of the obtained parameter sets again produce fits of similar quality to that of \citet{Fabretti13}, which we demonstrate in Section \ref{CalResults}.\ref{DataComparison}.

\subsection{Comparison of Calibrated Model and Empirical Data \label{DataComparison}}

It was mentioned in the preceding sections that while no unique parameter set could be established, all of the experiments converged to different parameter sets with similar objective function values producing reasonable behavior when compared to the simulated data. In this section, we consider the best parameter set, as measured by the objective function, obtained in the Nelder-Mead simplex algorithm calibration experiments.

For this parameter set, we simulate $20$ price paths and obtain the $95\%$ confidence intervals for the estimates of the mean, standard deviation, kurtosis, and Hurst exponent of the simulated log prices and compare this to the empirical moments obtained from the transaction data.

\begin{table}[H]
\caption{Best Parameter Set Obtained Through Calibration \label{BestParams}}
\begin{tabular*}{\linewidth}{@{\extracolsep{\fill}}ccc}
\hline
Calibrated Parameter & Parameter Value \\ \hline
$\delta$ & $0.0733$ \\
$\lambda_0$ & $180$ \\
$C_\lambda$ & $33$ \\
$\Delta_S$ & $0.0328$ \\
$\alpha$ & $0.2129$ \\
$\mu$ & $0.0653$ \\ \hline
\end{tabular*}
\vspace{0.2cm}
\caption*{Best parameter set obtained through the implementation of Nelder-Mead simplex algorithm combined with the threshold accepting heuristic}
\end{table}

\vspace{-0.5cm}

\begin{table}[H]
\caption{Comparison of Empirical and Calibrated Moments \label{CalibratedMoments}}
\begin{tabular*}{\linewidth}{@{\extracolsep{\fill}}ccc}
\hline
Moment & Simulated Value & Empirical Value\\ \hline
$m_1$ & $[5.4628, 5.5137]$ & $5.5143$ \\
$m_2$ & $[0.0268, 0.0387]$ & $0.0300$ \\ 
$m_3$ & $[2.3716, 3.3788]$ & $1.2402$ \\
$m_4$ & $[0.5795, 0.5993]$ & $0.5658$\\ \hline
\end{tabular*}
\vspace{0.2cm}
\caption*{$95\%$ confidence intervals for the simulated moments obtained using the calibrated parameters compared to the empirically measured moments, with $m_1,...,m_4$ corresponding to the mean, standard deviation, kurtosis and generalized Hurst exponent respectively}
\end{table}

\vspace{-0.5cm}

As seen in Table \ref{CalibratedMoments} and also demonstrated in prior work \citep{Platt16}, we observe that the parameter sets obtained through the calibration procedure employed are indeed able to produce fits of similar quality to that of \citet{Fabretti13} for the mean and standard deviation, while slightly overestimating the Hurst exponent and severely overestimating the kurtosis. This is predominantly due to the fact that the weight matrix obtained through the method of \citet{Fabretti13} tends to assign very large weights to errors on the mean and standard deviation, while assigning very small weights to errors on kurtosis, leading calibration to either underestimate or overestimate kurtosis in the investigations of both \cite{Fabretti13} and \citet{Platt16}.

We provide a detailed discussion on possible shortcomings of the method of \citet{Fabretti13} in Section \ref{CalibrationMethodAssessment}.

Despite the previously mentioned flaws, the above seems to suggest that members of this class of models, namely ABMs of continuous double auction markets using realistic order matching processes, are indeed capable of reproducing a number of characteristics of empirically-sampled data to some degree and that the calibration methodology of \citet{Fabretti13} can indeed determine appropriate parameter sets, but that the parameters producing such fits are simply not unique. In the following subsections, we detail possible reasons for this behavior through an investigation of the behavior of the objective function.

\subsection{Parameter Analysis \label{ParamAnalysis}}

In an attempt to illustrate the behavior of the objective function with respect to changes in parameter values, we generate surfaces representing the objective function values corresponding to various parameter value pairs, as has been done by \citet{Gilli03} and \citet{Fabretti13}. We repeat the process previously employed in \citet{Platt16}, where we consider $1000$ points in a two-dimensional space generated using a two-dimensional Sobol sequence, corresponding to various possible values of a parameter pair. We then obtain the objective function value for each combination of the values in the parameter pair using $5$ Monte Carlo replications, resulting in an objective function surface. The base parameter set used for these experiments is presented in Table \ref{BestParams}.

\begin{figure}[H]

\centerline{
\includegraphics[width=0.9\linewidth]{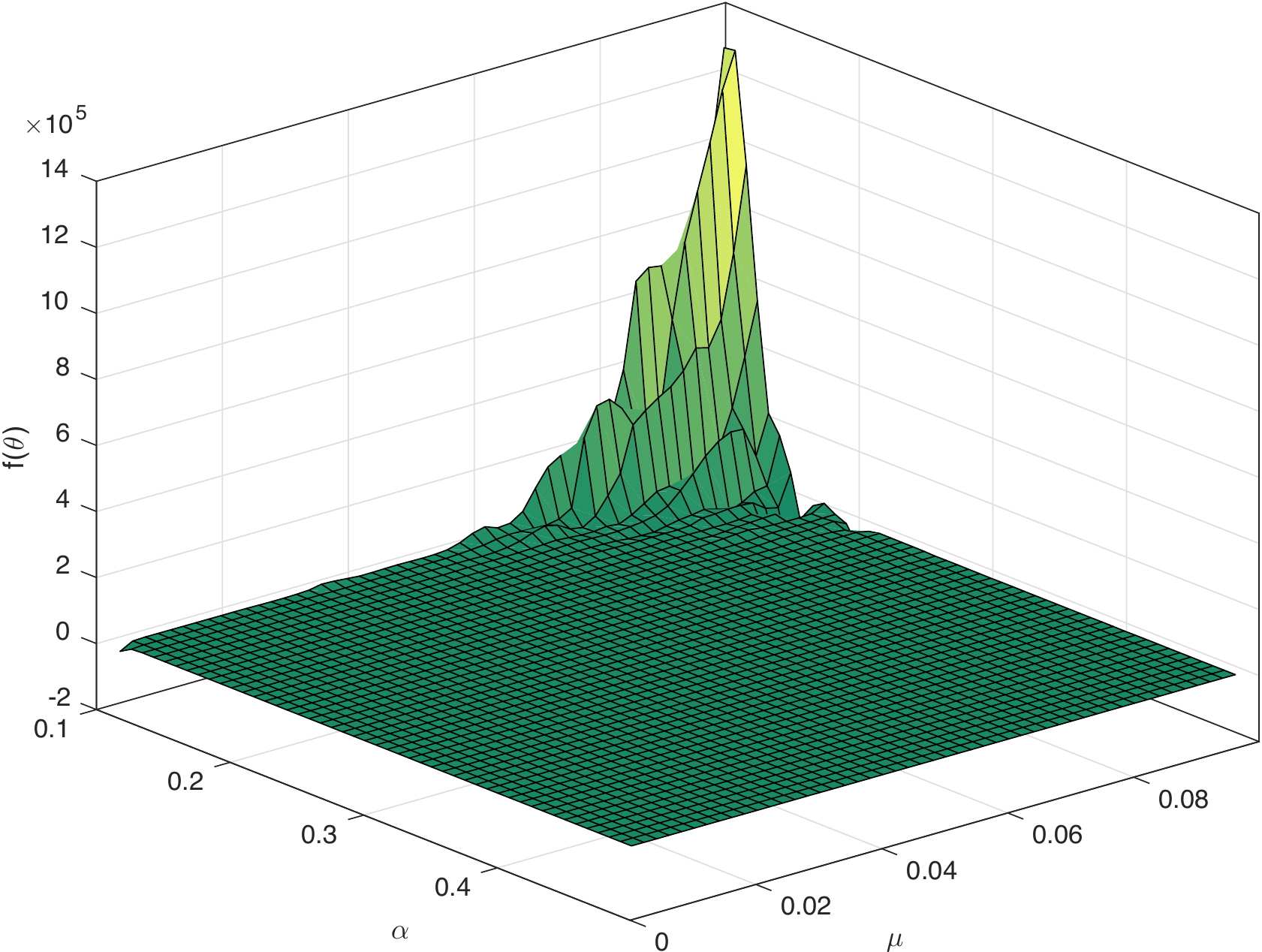}
}

\caption{Type 1 \citep{Platt16} parameter surface illustrating the effect of $\alpha$ and $\mu$ on the objective function \label{Surface_alpha_mu}}

\end{figure}

\begin{figure}[H]
\centerline{
\includegraphics[width=0.9\linewidth]{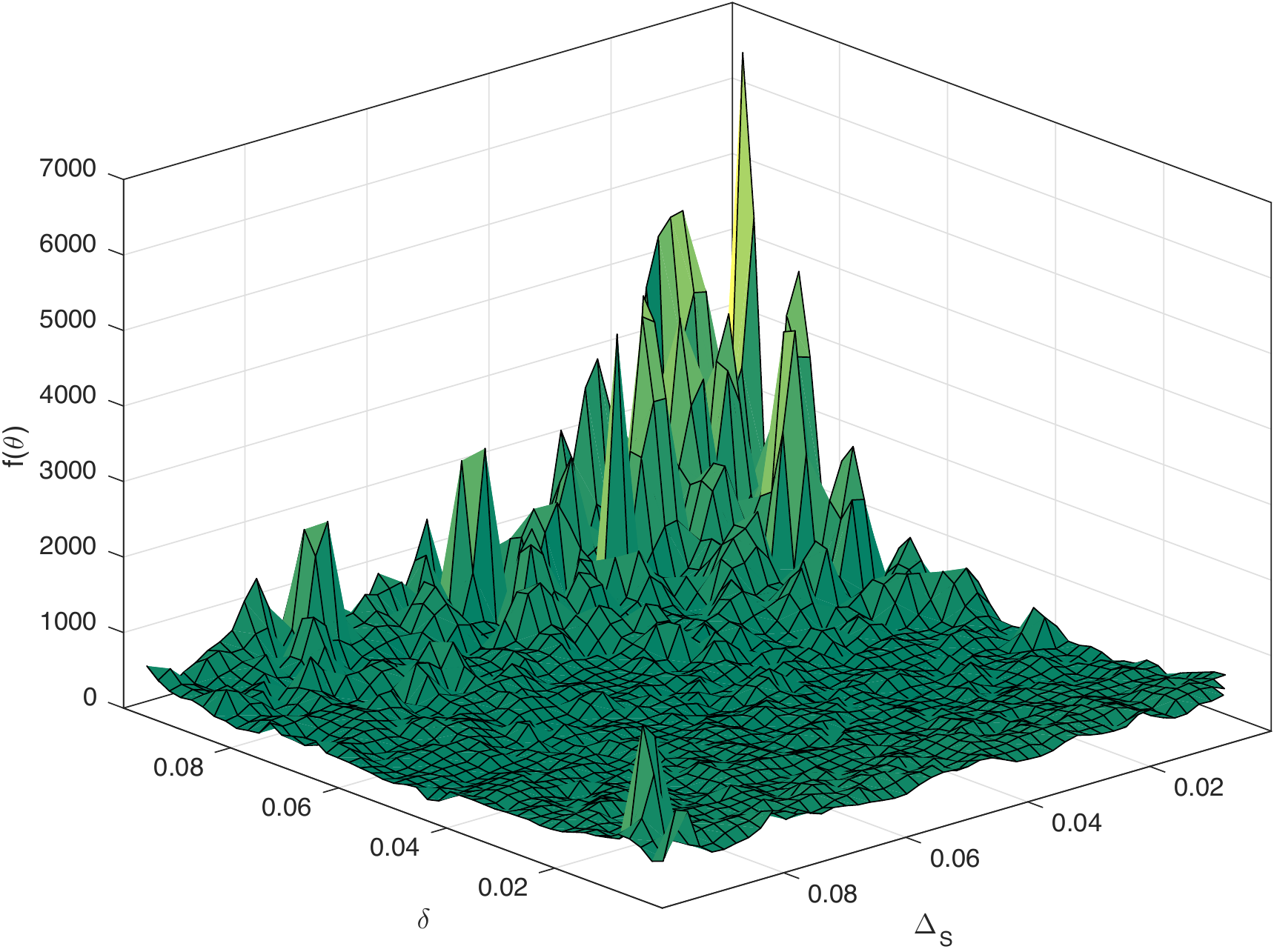}
}

\caption{Type 1 \citep{Platt16} parameter surface illustrating the effect of $\delta$ and $\Delta_S$ on the objective function \label{Surface_delta_delta_S}}

\end{figure}

As is clearly visible in Figures \ref{Surface_alpha_mu} and \ref{Surface_delta_delta_S}, we see that $\delta$, $\Delta_S$, $\alpha$ and $\mu$ all exhibit a similar effect on the objective function. While there are indeed certain regions that produce particularly bad fits to the data, the vast majority of the surfaces are relatively flat, indicating a very wide range of feasible parameters producing a reasonable fit to the data, with no clear way to distinguish an optimal global minimum. This is a similar insight to that expressed by \citet{Fabretti13}. 

This is the likely reason why parameter convergence was not observed, despite the fact that the objective function values were similar for each experiment, as a very large area of the parameter space produces very similar dynamics. The effect of these parameters on the dynamics of the obtained log price time series is thus not clearly defined, again indicative of parameter degeneracies in a similar vein to those found for the \citet{JacobLeal15} model in \citet{Platt16}.

\begin{figure}[H]
\centerline{
\includegraphics[width=0.9\linewidth]{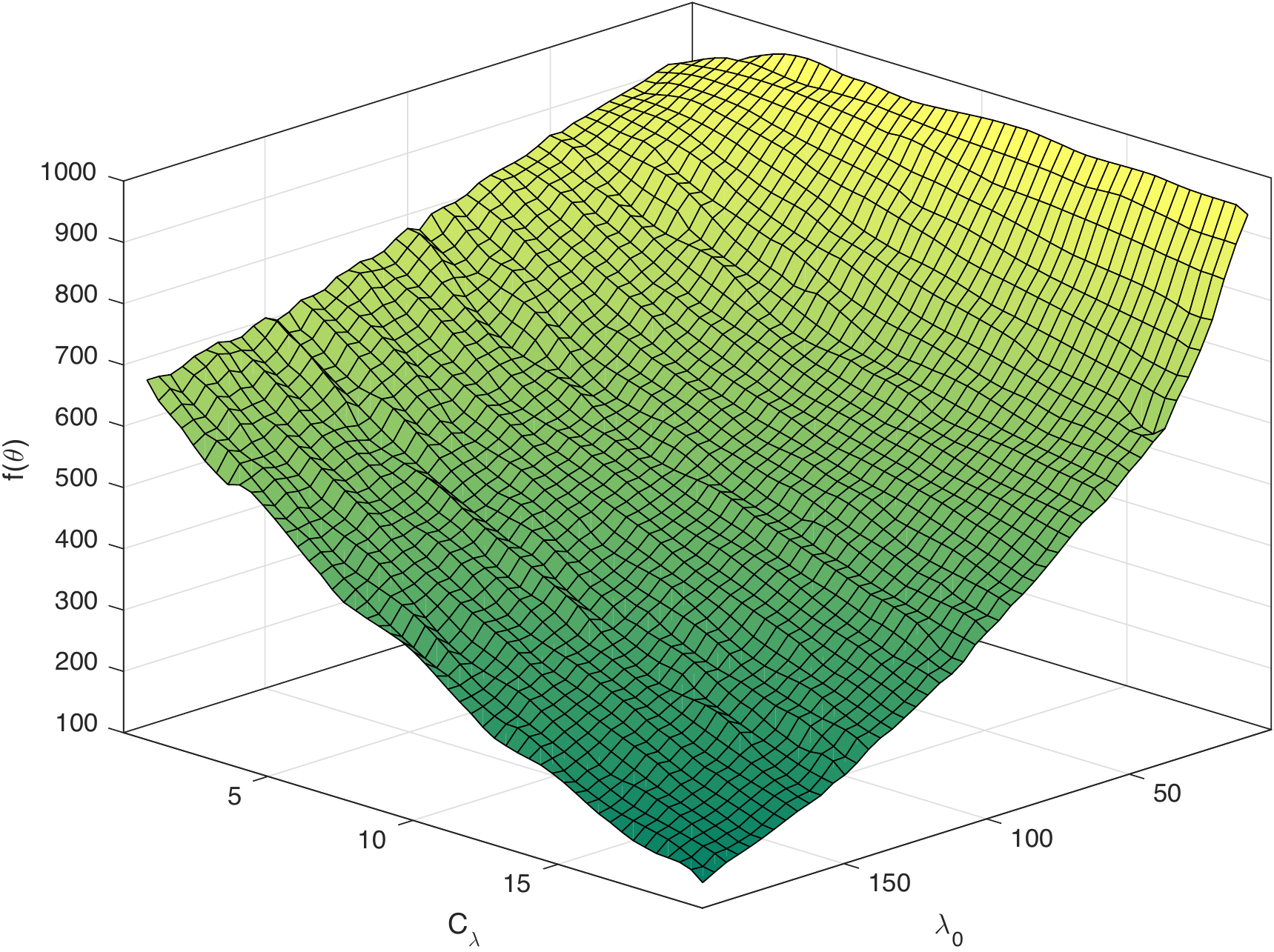}
}

\caption{Type 3 \citep{Platt16} parameter surface illustrating the effect of $\lambda_0$ and $C_\lambda$ on the objective function \label{Surface_lambda_0_C_lambda}}

\end{figure}

In contrast to this, however, Figure \ref{Surface_lambda_0_C_lambda} indicates that both $C_\lambda$ and $\lambda_0$ produce a very well-defined effect on the objective function. This is an interesting observation. In \citet{Platt16}, we observed a similar trend in the case of the \citet{JacobLeal15} model, with all parameter surfaces behaving in a haphazard manner, except for those dealing with parameters that drove the random walk which defined order prices in the order book. It is therefore not surprising that $C_\lambda$ and $\lambda_0$ happen to drive the order placement depth process, $\lambda(t)$, and hence order prices in this model.

In our previous investigation \citep{Platt16} we conjectured that this behavior could be attributed to the fact that the \citet{JacobLeal15} model set order prices according to the previous history of market prices only and not the current state of the order book. This may have reduced the effect of order size and order frequency effects on the obtained log price time series. This could, in turn, cause the associated parameters to become degenerate. It was believed that consideration of the order book dynamics during order placement, which would be more directly affected by the other parameters in the model, would lead to greater parameter stability.

Despite this, it now seems that this was not the case. We have found that even with such considerations in the \citet{Preis06} model, we have again produced similar findings, leading to a more unfortunate conclusion that increased mechanistic complexity may not enhance the ability of models to be open to credible calibration. 

In this context, one should consider the work of \citet{Bookstaber16}, which provides a number of significant mechanistic extensions to the \citet{Preis06} model, most notably the inclusion of heterogeneous decision cycles and new agent types, such as large firms. As with much of the existing ABM literature, stylized fact-centric validation procedures are employed exclusively, with the authors explicitly stating that such validation procedures are fairly rigorous. Our results indicate that this may in fact not be the case and that there is at least cause for some skepticism as to whether certain of the extensions to the model result in parameter degeneracies not detected by stylized fact-centric validation.

It is important to note that in the case of the \citet{JacobLeal15} model, a single parameter associated with one of the smooth objective function surfaces previously mentioned did show convergence, though neither $\lambda_0$ nor $C_\lambda$ have demonstrated this behavior in this investigation. 

This is most likely due to the fact that other parameters in the \citet{JacobLeal15} model never produced significant effects on the objective function in comparison to the single successfully calibrated parameter.

In the \citet{Preis06} model, however, there are regions where the objective function value can become very large for certain values of $\delta$, $\Delta_S$, $\alpha$ and $\mu$, whereas they typically have a very small effect on the objective function. This is because these regions typically represent unstable model dynamics, for example $\alpha$ close to $\mu$ and a high order cancelation rate, $\delta$, leading to an illiquid order book that is frequently depleted of orders, which in turn drives unrealistic and unstable simulated price dynamics.

These very high objective function values in these regions may have affected convergence in $\lambda_0$ and $C_\lambda$, since their effect would not be as dominant as it would be for typical parameter values, this dominance being the reason for the observed convergence in \citet{Platt16}. Rather than blindly accepting such reasoning, however, we rigorously demonstrate that such convergence is indeed possible in Section \ref{CalResults}.\ref{GA}.

It would therefore seem that regardless of the equations used to construct intraday ABMs of continuous double auction markets involving realistic order matching processes, the dynamics of the obtained market price time series are predominantly driven by the equations used to determine order prices in the model, with other parameters degenerating to a significant extent. 

This may suggest that prices are much more than just the aggregation of predefined mechanistic classes of agents \citep{Wilcox14}. 

\subsection{Verification of $\lambda_0$ and $C_\lambda$ Convergence \label{GA}}

In order to verify that $\lambda_0$ and $C_\lambda$ can be uniquely determined when the dynamics of the other parameters do not introduce noise into the calibration procedures, we consider a separate calibration experiment involving only $\lambda_0$ and $C_\lambda$. 

Rather than using the Nelder-Mead simplex algorithm as before, we now consider a genetic algorithm, since $3$ initial simplex vertices would be too small a parameter space to randomly initialize. 

We begin with a randomly initialized initial population of $100$ individuals, using the same parameter bounds as in the Nelder-Mead simplex calibration experiments, and iterate this population for $50$ generations, determined to be sufficient for convergence. We then repeat this process a total of $8$ times in order to obtain confidence intervals as in Section \ref{CalResults}.\ref{NM}. All other parameters are set according to the values in Table \ref{BestParams}.

\begin{table}[H]
\caption{Genetic Algorithm Calibration Results ($\lambda_0$, $C_\lambda$) \label{GAResults_1}}
\begin{tabular*}{\linewidth}{@{\extracolsep{\fill}}ccc}
\hline
Parameter & 95\% Conf Int & $\frac{s}{\sqrt{n}}$ \\ \hline
$\lambda_0$ & $[193.2596, 196.5567]$ & $0.6972$ \\
$C_\lambda$ & $[18.7165, 19.4306]$ & $0.1510$ \\ \hline
\end{tabular*}
\vspace{0.2cm}
\caption*{95\% confidence intervals for $\lambda_0$ and $C_\lambda$, obtained from 8 independent calibration experiments involving a genetic algorithm}
\end{table}

\vspace{-0.5cm}

Referring to Table \ref{GAResults_1}, we see that when only $\lambda_0$ and $C_\lambda$ are considered in a calibration experiment, we observe meaningful convergence in both parameters. The identified region of convergence is also consistent with the visible minimum in Figure \ref{Surface_lambda_0_C_lambda}.

\begin{table}[H]
\caption{Genetic Algorithm Calibration Results ($\delta$, $\Delta_S$, $\alpha$, $\mu$) \label{GAResults_2}}
\begin{tabular*}{\linewidth}{@{\extracolsep{\fill}}ccc}
\hline
Parameter & 95\% Conf Int & $\frac{s}{\sqrt{n}}$ \\ \hline
$\delta$ & $[0.0653, 0.0958]$ & $0.0064$ \\
$\Delta_S$ & $[0.0163, 0.0405]$ & $0.0051$ \\
$\alpha$ & $[0.1167, 0.1926]$ & $0.0160$ \\
$\mu$ & $[0.0608, 0.0924]$ & $0.0067$ \\ \hline
\end{tabular*}
\vspace{0.2cm}
\caption*{95\% confidence intervals for $\delta$, $\Delta_S$, $\alpha$ and $\mu$, obtained from 8 independent calibration experiments involving a genetic algorithm}
\end{table}

\vspace{-0.5cm}

Finally, we repeat the genetic algorithm experiment employed for parameters $\lambda_0$ and $C_\lambda$, but this time consider $\delta$, $\Delta_S$, $\alpha$ and $\mu$ as free parameters and as expected find that these parameters cannot be uniquely identified using the genetic algorithm, as shown in Table \ref{GAResults_2}.

\section{Realistic Order Matching Procedures and Price Dynamics}

In both our current and previous investigations \citep{Platt16}, it appears that certain features of the microstructure of a market, such as order book depth and the associated dynamics of the prices of limit orders, can be reliably probed by the calibration of intraday ABMs to transaction data. In contrast to this, however, it seems that the parameters of these models associated with the frequency at which traders submit orders, the specific number of trader agents participating in a particular market and various other phenomena that are rooted in agent behavior tend to be difficult to infer with significant confidence.

This naturally leads to a discussion on the effect of various processes within an intraday ABM on the simulated price time series when this time series is generated using realistic order matching processes. \citet{Bouchaud08} argue that prices in real markets are driven by two main phenomena, the response of prices to individual orders and the flow of orders arriving in the market. We therefore investigate how this may apply to a market simulated using an intraday ABM.

It is evident that the processes which determine order prices within an intraday ABM directly influence the first of these phenomena in a well-defined way. These processes directly determine the prices of limit orders within the LOB and these order prices define the mid prices or trade prices (depending on the model in question) that will eventually be set as market prices. This likely explains the fairly well-defined effect of the parameters influencing order prices on both the simulated price time series and objective function in the case of the \citet{JacobLeal15} and \citet{Preis06} models. 

The second mechanism, namely the order flow, requires more careful consideration. A key question to consider is whether the \citet{Preis06} model does in fact produce a realistic, nontrivial order flow. 

A well-known property of order flow in real financial markets is long memory \citep{Bouchaud08}. This phenomenon can be observed by plotting the autocorrelation function of the time series of trade signs for a particular market over some period (+1 for buyer initiated trades and -1 for seller initiated trades), which typically demonstrates a slow decay in autocorrelation as the number of lags is increased in most empirical investigations \citep{Kuroda11}.

\begin{figure}[H]
\centerline{
\includegraphics[width=1\linewidth]{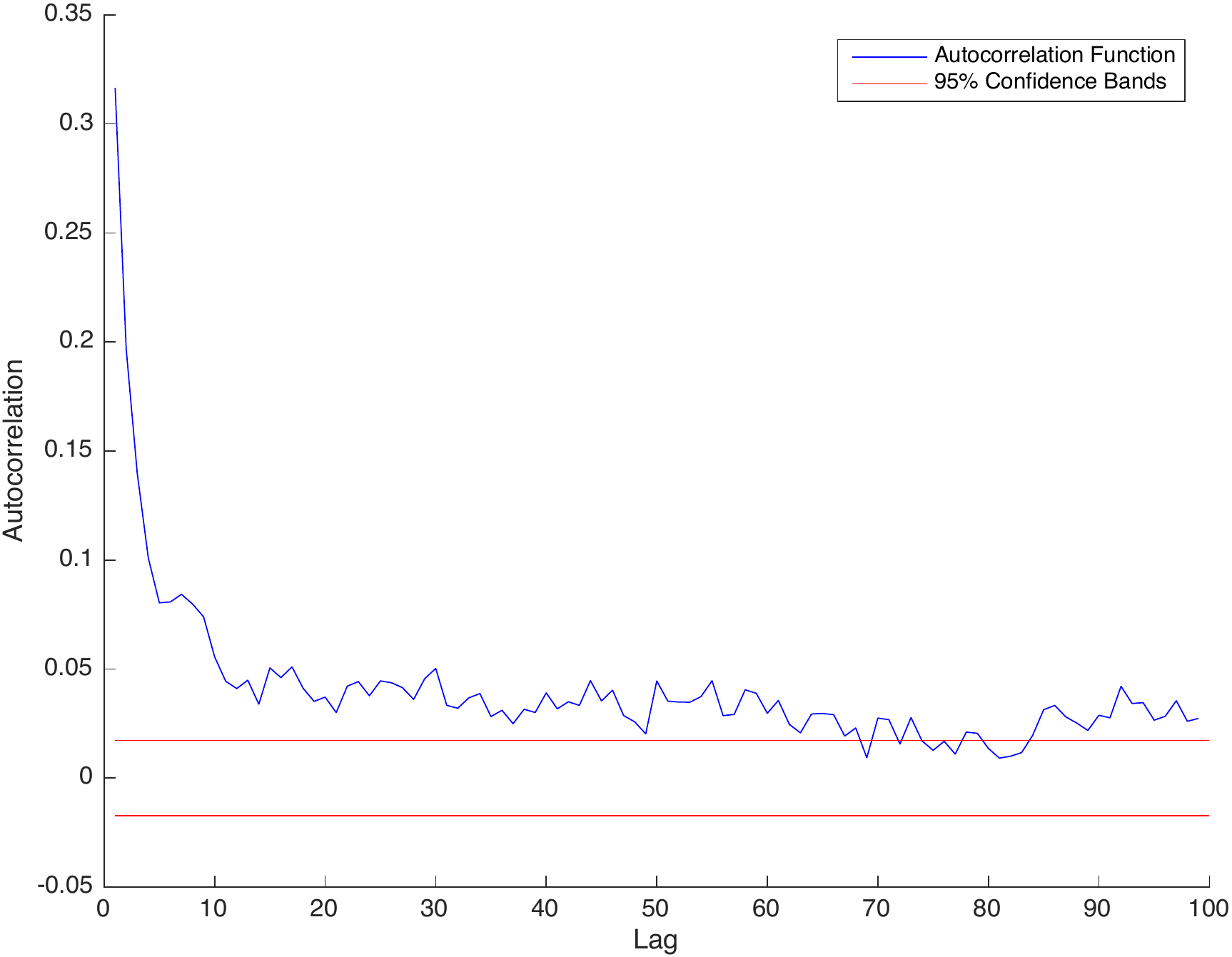}
}

\caption{Autocorrelation function for trade signs inferred from our calibration dataset \label{MeasuredOrderFlow}}

\end{figure}

Using the measured data and following the method of \citet{Lee91}, which was also applied to similar datasets in \citet{Harvey17}, we classified trades as either buyer or seller initiated and computed the associated trade sign autocorrelation function. This autocorrelation function is presented in Figure \ref{MeasuredOrderFlow} and confirms that long memory dynamics are captured in the order flow associated with the data to which the model was calibrated.

\begin{table}[H]
\centering
\caption{Default Model Parameter Set \label{DefaultParameters}}

\begin{tabular*}{\linewidth}{@{\extracolsep{\fill}}cccc}
\hline
& Parameter & Default Value \\ \hline
& $\delta$ & $0.0250$ \\
& $\lambda_0$ & $100$ \\
& $C_\lambda$ & $10$ \\
& $\Delta_S$ & $0.0010$ \\
& $\alpha$ & $0.1500$ \\
& $\mu$ & $0.0250$ \\ \hline
\end{tabular*}

\vspace{0.2cm}
\caption*{Default model parameter set presented in \citet{Preis06} and \citet{Preis07}}

\end{table}

\vspace{-0.5cm}

\begin{figure}[H]
\centerline{
\includegraphics[width=1\linewidth]{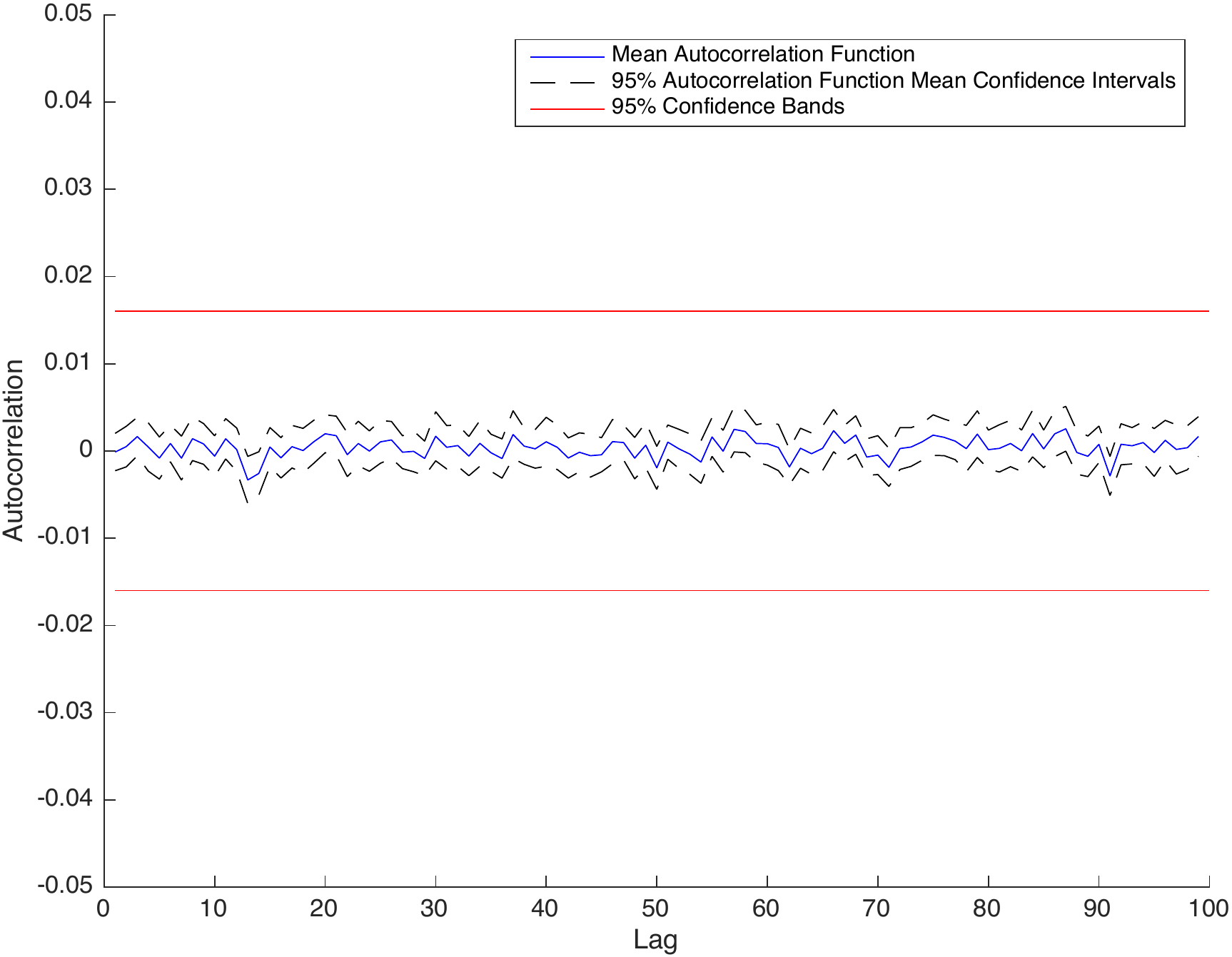}
}

\caption{Autocorrelation function for trade signs generated over $50$ \citet{Preis06} model simulations initialized with the parameter set presented in Table \ref{DefaultParameters} \label{TradeSignACFDef}}

\end{figure}

Figure \ref{TradeSignACFDef} shows the autocorrelation function for trade signs generated by the \citet{Preis06} model, initialized with the default parameters described by \citet{Preis06}, which we present in Table \ref{DefaultParameters}. We see that the characteristic slow decay in autocorrelation associated with long memory is absent, indicative that such dynamics are not captured by the model when initialized with the default parameters of \citet{Preis06}.

\begin{figure}[H]
\centerline{
\includegraphics[width=1\linewidth]{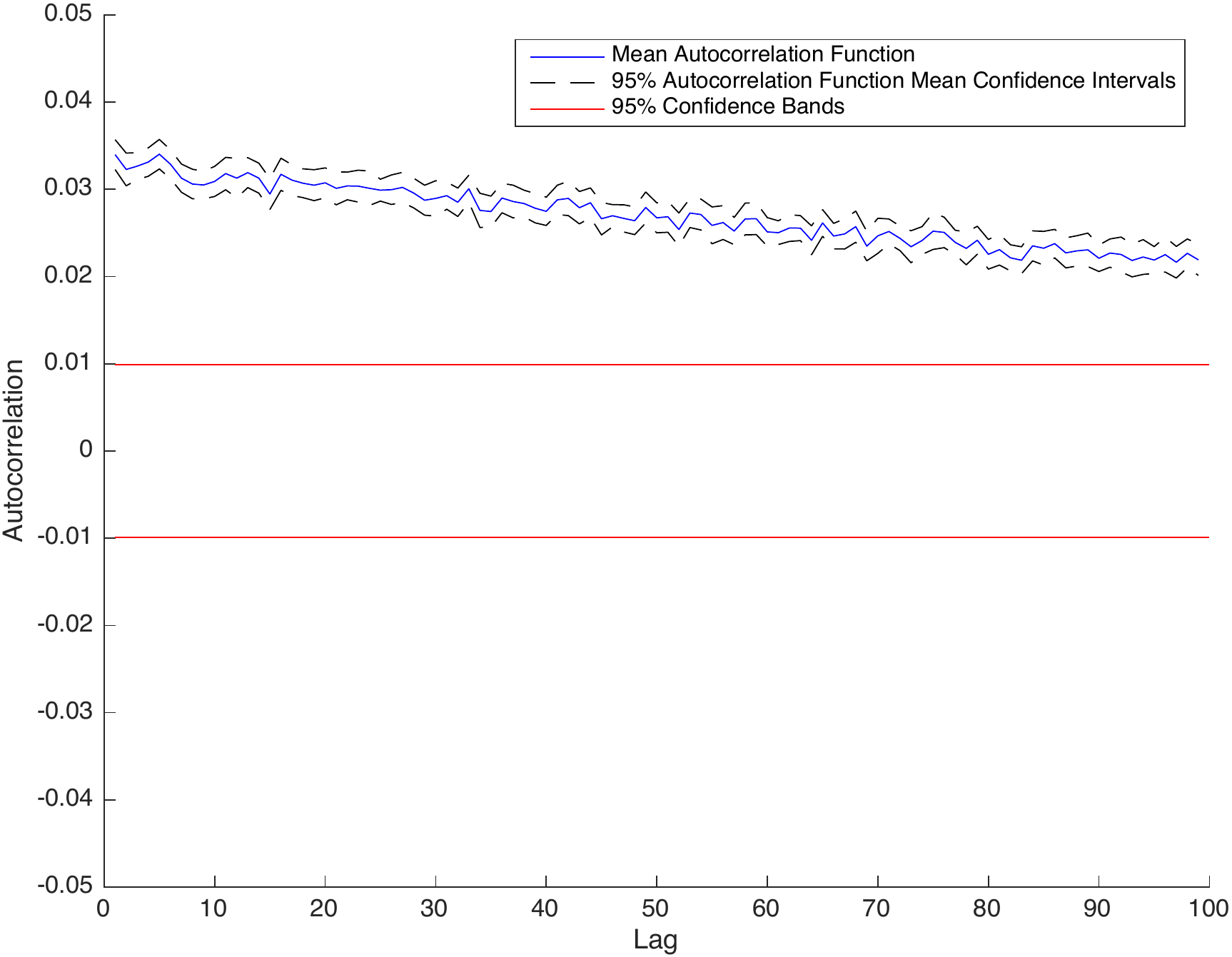}
}

\caption{Autocorrelation function for trade signs generated over $50$ \citet{Preis06} model simulations initialized with the parameter set presented in Table \ref{BestParams} \label{TradeSignACFCal}}

\end{figure}

Referring to Figure \ref{TradeSignACFCal}, where we repeat this exercise using the calibrated parameters presented in Table \ref{BestParams}, we see that the slow decay in autocorrelation associated with long memory processes is now present, indicating that the calibration process has determined a parameter combination that results in a more realistic order flow than the model's default parameters.

On closer inspection, it was found that shifting the values of individual parameters in Table \ref{BestParams}, with the exception of $\Delta_S$, and repeating this exercise did not cause these more realistic long memory dynamics to disappear. When shifting $\Delta_S$, however, it was found that smaller values of this parameter resulted in long memory dynamics no longer being present in the order flow.

This is in fact consistent with the surface presented in Figure \ref{Surface_delta_delta_S}: 

For the $\delta$ value presented in Table $\ref{BestParams}$, approximately $0.07$, we see that smaller $\Delta_S$ values tend to result in poorer objective function values, whereas values of $\Delta_S$ roughly larger than $0.03$ tend to result in fairly similar and relatively low objective function values. Not surprisingly, this region of lower objective function values in fact corresponds relatively well to those in which long memory was observed in the simulated order flow.

Therefore, $\Delta_S$ does indeed exert some effect on the order flow within the model, but a very wide range of possible values exists for $\Delta_S$ that produces long memory dynamics. This is reflected in the effect of $\Delta_S$ on the obtained price time series, which is itself reflected in the effect of $\Delta_S$ on the objective function, leading to a large region where no unique minimum can be determined. While the simulated order flow obtained through calibration is by no means identical to that observed in the data, likely due to model limitations, calibration has resulted in an attempt to capture these dynamics.

In contrast to $C_\lambda$, $\lambda_0$ and $\Delta_S$, parameters $\alpha$, $\mu$ and $\delta$ relate to agent behaviors and thus do not directly affect the individual orders or order flow within the model. The fact that altering these parameters did not affect the presence of long memory in the simulated order flow provides evidence of the second aspect of the preceding statement. This implies that these parameters do not directly relate to the two processes conjectured by \citet{Bouchaud08} to be the driving forces behind prices in real markets and this seems to be mirrored in the realistic matching processes employed in intraday ABMs, where we observed that the effect of these parameters on the simulated price time series was not well-defined.

We therefore conjecture that the encountered parameter degeneracies emerge from the application of realistic matching processes within intraday ABMs. Parameters which exert well-defined effects on either order flow or individual orders themselves produce well-defined effects on the obtained simulated price time series and are therefore amenable to calibration. These parameters tend to be rooted in the microstructure of the market and the nature of the LOB. In contrast to this, parameters which relate more to agent behavior and do not tend to have a well-defined or direct effect on individual order dynamics or order flow tend to produce poorly-defined effects on the obtained price time series, resulting in calibration difficulties.

This may explain why the calibration of closed-form, daily ABMs proved more successful in past experiments, such as those of \citet{Fabretti13}. In using a closed-form solution to determine market prices, one directly enforces a well-defined, albeit simplified, relationship between parameter values and the obtained market price time series. In contrast to this, the realistic matching processes in intraday models seem to be instead driven by order flow and the nature of individual orders, leading parameters which do not directly determine these dynamics in a well-defined way to produce haphazard dynamics.

The recovery of trade sign autocorrelation through model calibration leads us to a discussion of the latent order book. This is a key potential source of top-down causation \citep{Wilcox14} that may render the type of ABM considered here as being fatally incomplete. 

As suggested by \citet{Bonart15}, the LOB itself only represents a fraction of the actual liquidity in a particular market. In real financial markets, liquidity providers attempt to hide their intentions and thus tend to avoid the placement of very large limit orders. This implies that liquidity takers attempting to acquire or liquidate large numbers of shares cannot do so through the placement of a single order and thus break a single parent order into a number of child orders, resulting in the observed trade sign autocorrelation \citep{Toth11}. The true intentions of liquidity providers and the true nature of supply and demand are therefore not captured in the order book itself, but rather in a fictitious order book known as the latent order book. The purpose of trading that drives the latent liquidity can operate on very different, and possibly much longer, time scales than those inherent in the short term dynamics of the order book \citep{Bouchaud08}.

Therefore, it may be worthwhile to consider the calibration of reaction-diffusion models of the latent order book to similar data to that employed in this investigation in future work. Possible candidate models include the work of \citet{Mastromatteo14a}, \citet{Mastromatteo14b} and \citet{Gao14}.

It is clear that the ABM considered, though not explicitly considering the latent order book, has been able to reproduce some of the features of financial markets thought to originate from it, suggesting that there may be some hidden explanatory potential embedded within ABMs of continuous double auction markets with regard to the underlying microstructure of the market. The ability to uniquely determine the parameters associated with limit order price generation processes in both the \citet{JacobLeal15} and \citet{Preis06} models provides further evidence of this possibility. 

Therefore, while certain parameters could not be uniquely determined in our calibration experiments, this does not imply that such models do not possess an ability to probe market microstructure.

If dynamics relating to possible sources of top-down causation \citep{Wilcox14}, such as those resulting from the latent order book \citep{Bouchaud08, Bonart15}, were not readily recoverable through calibration, there would be concerns that a given model could not be used to faithfully represent the structure of the market being modeled. We have been able to provide an example of the use of calibration to determine parameters that recover reasonable order flow autocorrelation without a priori parameter tuning. 

This does not resolve the concerns that the feedbacks between order flow and the state of the LOB may be dominant features of real intraday markets and that the behaviors of different agent classes cannot be robustly resolved through calibration. This leads to the conclusion that overly complex and mechanistic intraday ABMs and market clearing specifications may be counterproductive in the search for parsimonious, yet useful model representations.

The derivation of models from fundamental assumptions about the intelligent behaviors of market participants often results in models that are difficult to test \citep{Ziliak04}, since a large variety of auxiliary assumptions are often necessary to make model estimation meaningful. These assumptions may enhance the formal acceptability of a particular method or approach, but often lead to little new insight into how best to model, explain or predict phenomenology arising from real markets. This can be seen as resulting in a situation where a model seems to have passed a variety of tests but can still present model behaviors that can be recovered using alternative explanations \citep{Farmer05, Ziliak04}.

As stated in \citet{Farmer05} and alluded to in \citet{Smith03}, it is conceivable to have a successful model that does not directly depend on the nuances of agent behaviors, even when it is known that the real agents engage in sophisticated behaviors. The fact that our calibration attempts point to a similar conclusion, though from a very different departure point, could be telling us something important about the relationship between rational effects and stochastic effects -- particularly where institutions can strongly shape behavior \citep{Farmer05}. This points to the widely held view \citep{Bouchaud08} that it is indeed the combination of order-flow and price specification that drive prices in financial markets. As stated in the \citet{Farmer05}, this does not answer why order flow varies as it does but seems to point at the situation where ABMs of intraday double-auction markets can be sufficiently well described by a reduced set of parameters, parameters that are directly related to order flow and order price specification, independently of complex agent behaviors.

\section{Relevant Caveats \label{CalibrationMethodAssessment}}

It is important to note that our investigation does not claim to be definitive or exhaustive, but rather aims to encourage new lines of investigation that further evaluate the rigor of current research paradigms, such as the exclusive use of stylized fact-centric validation and the continued increase of the mechanistic complexity of financial ABMs without compelling model calibration.

While a substantial part of this manuscript has been devoted to the discussion of possible flaws relating to current approaches to intraday financial ABM construction, it is also important to identify possible flaws associated with the calibration methodology itself as a relevant caveat to this investigation, which we alluded to in \citet{Platt16}.

Although the method of \citet{Fabretti13} is well-motivated for the most part and based upon established literature, it does involve the inversion of near-singular matrices. We believe, however, that because we are able to obtain smooth objective function surfaces for certain parameters ($\lambda_0$, $C_\lambda$) in both our current and previous investigations \citep{Platt16}, any noisy effects observed in the constructed objective function surfaces most likely stem from model as opposed to calibration method-related concerns. 

In addition to this and as previously discussed, the weight matrices in both our investigation and that of \citet{Fabretti13} assign significantly larger weights to errors on the mean and standard deviation in comparison to other moments. Given that characteristics such as kurtosis and the Hurst exponent are relatively important features of financial return time series and since there may be many different distributions that could be parameterized by the same mean and standard deviation, this is also an undesirable feature of the method.

Therefore, it is recommended that future investigations involving the method of \citet{Fabretti13} or similar methods make attempts to address these particular problems to improve numerical stability and the veracity of any results obtained using such methods. In achieving such ends, one might also consider the replacement of the constructed objective function with an alternative benchmark, such as the information theoretic criterion introduced by \citet{Lamperti15}. 

While we have identified a number of relevant methodological concerns, we believe that the identification of relatively precise values for $\lambda_0$ and $C_\lambda$ as well as the ability of the method to find parameter sets that recover order flow correlations are sufficient to argue that the method of \citet{Fabretti13} is able to provide valuable insights in this context.

\section{Conclusion}

In both our current and previous investigations \citep{Platt16}, we have observed that there is some evidence indicating that intraday ABMs of continuous double auction markets employing realistic order matching procedures tend to produce price time series with dynamics predominantly driven by the processes or equations defining order prices within them, with the remaining parameters and procedures tending to produce either relatively insignificant or poorly behaved effects on the same price time series. This is indicative of parameter degeneracies and is in spite of the fact that these models tend to reproduce the empirically-observed stylized facts of financial return time series relatively well.

We conjecture that this is likely due to the fact that like in real financial markets, simulated market prices in intraday ABMs employing realistic matching processes are driven by responses to individual orders and the nature of order flow and therefore only parameters associated with processes which affect these phenomena in well-defined ways produce well-defined and significant effects on the market price time series.

This suggests that parameters related to these phenomena, such as the nature of individual orders in a market, can indeed be probed by intraday ABMs, but that parameters rooted in agent behavior are difficult to calibrate due to the fact that they do not exert well-behaved effects on the simulated price time series when it is generated using realistic matching processes.

Traditional stylized fact-centric validation seems unable to detect these potential problems, suggesting that such methods of validation are simply not sufficient for the validation of intraday ABMs employing realistic matching procedures. Therefore, simply increasing the mechanistic complexity of agents in intraday ABMs such that empirically-observed return time series stylized facts are reproduced may lead to flawed insights, simply because the relationship between the simulated price time series and agent behaviors has to be well-defined.

We thus argue that calibration and stylized-fact replication should both be considered when developing such ABMs, in order to ensure a replication of reasonable model behavior, while still ensuring that parameters produce the intended effect of the model designer on the simulated price time series.

This move towards more rigorous validation would be an important paradigm shift, since our existing calibration experiments have already shown evidence of dynamics, such as order flow correlation, not being sufficiently well-captured in the models themselves, but emerging as important concerns when attempting to calibrate such models to data.

\section{Acknowledgements}
T.J. Gebbie acknowledges the financial support of the National Research Foundation (NRF) of South Africa (grant number 87830). D.F. Platt acknowledges the financial support of the University of the Witwatersrand, Johannesburg and NRF of South Africa. The conclusions herein are due to the authors and the NRF accepts no liability in this regard.

\end{multicols}


\begin{thebibliography}{99}

\bibitem[Barde(2016)]{Barde16}
Barde S (2016) Direct calibration and comparison of agent-based herding models of financial markets. J Econ Dyn Control 73:329-353

\bibitem[Bonart(2015)]{Bonart15}
Bonart J (2015) Mathematical aspects of delayed market clearing in order driven markets and its applications to non-Markovian price impact and optimal execution. SSRN:2659092

\bibitem[Bookstaber et al.(2016)]{Bookstaber16}
Bookstaber R, Foley MD, Tivnan BF (2016) Toward an understanding of market resilience: market liquidity and heterogeneity in the investor decision cycle. J Econ Interact Coord 11:205-227

\bibitem[Bouchaud et al.(2008)]{Bouchaud08}
Bouchaud JP, Farmer JD, Lillo F (2008) How markets slowly digest changes in supply and demand. arXiv:0809.0822.

\bibitem[Cartea et al.(2015)]{Cartea15}
Cartea A, Jaimungal S, Penalva J (2015) Algorithmic and High-Frequency Trading. Cambridge University Press, Cambridge

\bibitem[Chiarella and Iori(2002)]{Chiarella02}
Chiarella C, Iori G (2002) A simulation analysis of the microstructure of double auction markets. Quant Financ 2:346-353

\bibitem[Chiarella et al.(2009)]{Chiarella09}
Chiarella C, Iori G, Perello J (2009) The impact of heterogeneous trading rules on the limit order book and order flows. J Econ Dyn Control 33:525-537

\bibitem[Cont(2001)]{Cont01}
Cont R (2001) Empirical properties of asset returns: stylized facts and statistical issues. Quant Financ 1:223-236

\bibitem[Easley et al.(2012)]{Easley12}
Easley D, Lopez de Prado M, O'Hara M (2012) The Volume Clock: Insights into the High-Frequency Paradigm. J Portfolio Manage 39:19-29

\bibitem[Fabretti(2013)]{Fabretti13}
Fabretti A (2013) On the problem of calibrating an agent-based model for financial markets. J Econ Interact Coord 8:277-293

\bibitem[Farmer and Joshi(2002)]{Farmer02}
Farmer JD, Joshi S (2002) The price dynamics of common trading strategies. J Econ Behav Organ 49:149-171

\bibitem[Farmer et al.(2005)]{Farmer05}
Farmer JD, Patelli P, Zovko II (2005) The predictive power of zero intelligence in financial markets. Proc Natl Acad Sci USA 102(6):2254-2259

\bibitem[Gao and Deng(2014)]{Gao14}
Gao X, Deng SJ (2014) Hydrodynamic limit of order book dynamics. arXiv:1411.7502

\bibitem[Gilli and Winker(2003)]{Gilli03}
Gilli M, Winker P (2003) A global optimization heuristic for estimating agent based models. Comput Stat Data Anal 42:299-312

\bibitem[Hamill and Gilbert(2016)]{Hamill16}
Hamill L, Gilbert N (2016) Agent-Based Modelling in Economics. John Wiley \& Sons, Chichester

\bibitem[Harvey et al.(2017)]{Harvey17}
Harvey M, Hendricks D, Gebbie T, Wilcox D (2017) Deviations in expected price impact for small transaction volumes under fee restructuring. Physica A 471:416-426

\bibitem[Jacob Leal et al.(2015)]{JacobLeal15}
Jacob Leal S, Napoletano M, Roventini A, Fagiolo G (2015) Rock Around the Clock: An Agent-Based Model of Low- and High-frequency Trading. J Evol Econ 25:1-25

\bibitem[Kirman(1991)]{Kirman91}
Kirman A (1991) Epidemics of opinion and speculative bubbles in financial markets. In: Taylor M (ed) Money and Financial Markets. Blackwell, Oxford, pp 354-368

\bibitem[Kukacka and Barunik (2016)]{Kukacka16}
Kukacka J, Barunik J (2016) Estimation of Financial Agent-Based Models with Simulated Maximum Likelihood. SSRN:2783663

\bibitem[Kuroda et al.(2011)]{Kuroda11}
Kuroda K, Maskawa J, Murai J (2011) Stock price process and long memory in trade signs. Adv Math Econ 14:69-92

\bibitem[Lamperti(2015)]{Lamperti15}
Lamperti F (2015) An Information Theoretic Criterion for Empirical Validation of Time Series Models. LEM Papers Series, Laboratory of Economics and Management, Sant'Anna School of Advanced Studies, 2-2015

\bibitem[LeBaron(2005)]{LeBaron05}
LeBaron B (2005) Agent-based Computational Finance. In: Judd KL, Tesfatsion L (eds) The Handbook of Computational Economics, Vol. 2. Elsevier, Amsterdam, pp 1187-1233

\bibitem[Lee and Ready(1991)]{Lee91}
Lee CMC, Ready MJ (1991) Inferring Trade Direction from Intraday Data. J Finance 46(2):733-746

\bibitem[Mandes(2015)]{Mandes15}
Mandes A (2015) Order placement in a continuous double auction agent based model. Algorithmic Finance 4(3-4):105-125

\bibitem[Mastromatteo et al.(2014a)]{Mastromatteo14a}
Mastromatteo I, T\'oth B, Bouchaud JP (2014a) Anomalous Impact in Reaction-Diffusion Financial Models. Phys Rev Lett 113:268701

\bibitem[Mastromatteo et al.(2014b)]{Mastromatteo14b}
Mastromatteo I, T\'oth B, Bouchaud JP (2014b) Agent-based models for latent liquidity and concave price impact. Phys Rev E 89:042805

\bibitem[Panayi et al.(2013)]{Panayi13}
Panayi E, Harman M, Wetherilt A (2013) Agent-based modelling of stock markets using existing order book data. In: Giardini F, Amblard F (eds) Multi-Agent-Based Simulation XIII. Springer-Verlag, Berlin, pp 101-114

\bibitem[Platt and Gebbie(2016)]{Platt16}
Platt DF, Gebbie TJ (2016) The Problem of Calibrating an Agent-Based Model of High-Frequency Trading. arXiv:1606.01495

\bibitem[Preis et al.(2006)]{Preis06}
Preis T, Golke S, Paul W, Schneider (2006) Multi-agent-based Order Book Model of financial markets. Europhys Lett 75(3):510-516

\bibitem[Preis et al.(2007)]{Preis07}
Preis T, Golke S, Paul W, Schneider (2007) Statistical analysis of financial returns for a multiagent order book model of asset trading. Phys Rev E 76:016108

\bibitem[Smith et al.(2003)]{Smith03}
Smith E, Farmer JD, Gillemot L, Krishnamurthy S (2003) Statistical theory of the continuous double auction. Quant Financ 3:481-514

\bibitem[Thomson Reuters(2016)]{Reuters16}
Thomson Reuters (2016) Thomson Reuters Tick History. https://tickhistory.thomsonreuters.com. Accessed 23 May 2016

\bibitem[T\'oth et al.(2011)]{Toth11}
T\'oth B, Lemp\'eri\`ere Y, Deremble C, de Lataillade J, Kockelkoren J, Bouchaud JP (2011) Anomalous Price Impact and the Critical Nature of Liquidity in Financial Markets. Phys Rev X 1:021006

\bibitem[Wilcox and Gebbie(2014)]{Wilcox14}
Wilcox D, Gebbie T (2014) Hierarchical causality in financial economics. SSRN:2544327

\bibitem[Winker et al.(2007)]{Winker07}
Winker P, Gilli M, Jeleskovic V (2007) An Objective Function for Simulation Based Inference on Exchange Rate Data. SSRN:964131

\bibitem[Ziliak and McCloskey(2004)]{Ziliak04}
Ziliak ST, McCloskey DN (2004) Size matters: the standard error of regressions in the American Economic Review. J Socio Econ 33(5):527-546

\end{thebibliography}
\end{document}